\documentclass[aps,pra,twocolumn,showpacs,nofootinbib]{revtex4-1}
\usepackage{multirow}
\usepackage{graphicx}
\usepackage{amsmath}
\usepackage{dcolumn}

\begin{document}
\title{Long-range Rydberg-Rydberg interactions in calcium, strontium and ytterbium}
\author{C L Vaillant, M P A Jones and R M Potvliege}
\affiliation{Department of Physics, Durham University, South Road, Durham DH1 3LE}
\email{c.l.j.j.vaillant@durham.ac.uk}
\begin{abstract}
Long-range dipole-dipole and quadrupole-quadrupole interactions between pairs of Rydberg atoms are calculated perturbatively for calcium, strontium and ytterbium within the Coulomb approximation. Quantum defects, obtained by fitting existing laser spectroscopic data, are provided for all $S$, $P$, $D$ and $F$ series of strontium and for the $^3P_2$ series of calcium. The results show qualitative differences with the alkali metal atoms, including isotropically attractive interactions of the strontium $^1S_0$ states and a greater rarity of F\"orster resonances. Only two such resonances are identified, both in triplet series of strontium. The angular dependence of the long range interaction is briefly discussed.
\end{abstract}

\maketitle
\section{Introduction}
In the context of laser cooling, there has been a resurgence of interest in the strong interactions between Rydberg atoms.
A key development was the concept of the dipole blockade, where the strong interaction between Rydberg atoms leads to the excitation of collective states with a single Rydberg excitation shared between many atoms \cite{Lukin2001}. Applications of the dipole blockade include quantum information \cite{Saffman2010}, where two-qubit gates have already been demonstrated \cite{Wilk2010,Isenhower2010}, co-operative nonlinear optics \cite{Pritchard2010} and the physics of strongly correlated systems \cite{Cinti2010,Pohl2010,Schachenmayer2010}. Other important areas of study in cold Rydberg gases include the formation of long-range molecules \cite{Bendkowsky2009,Overstreet2009}, and the interplay between Rydberg gases and ultracold plasmas \cite{Robinson2000}.

Central to all of these applications is a detailed understanding of the long-range interactions between Rydberg atoms. As most experiments so far have been carried out using the alkali metals (Rb,Cs), theoretical work has concentrated on these elements \cite{Walker2008,Singer2005,Samboy2011,Reinhard2007,Schwettmann2006,Stanojevic2006}. Both perturbative calculations of the quadrupole-quadrupole and dipole-dipole interactions \cite{Walker2008,Singer2005, Reinhard2007} and detailed non-perturbative calculations of the dipole-dipole interactions \cite{Samboy2011,Stanojevic2006,Schwettmann2006} have been performed.

Atoms with two valence electrons offer a new approach, as the inner valence electron provides an additional way to probe and manipulate Rydberg atoms. Recent experiments have shown that the inner electron can be used as a fast, state-selective probe of the interactions in a cold Rydberg gas \cite{Millen2010,Millen2011}. The polarizability of the additional electron also enables tight, magic-wavelength traps for Rydberg atoms \cite{Mukherjee2011}. Rydberg states of Sr and Yb have also been proposed for high-precision measurements of the black-body shift in optical frequency standards \cite{Ovsiannikov2011}.

Preliminary calculations of the interactions for strontium revealed interesting features that differ from the alkali metals, such as the possibility of an isotropic, attractive interaction potential \cite{Mukherjee2011}. The goal of this work is to carry out a systematic study of the interactions between Rydberg atoms in the most commonly laser-cooled two-electron systems, i.e. Ca, Sr and Yb. In section \ref{Methods} we present the background theory for the calculations. The $C_5$ and $C_6$ coefficients governing the strength of the long-range interactions are expressed in terms of angular factors and radial matrix elements. The latter are evaluated using the Coulomb approximation, which requires accurate knowledge of the Rydberg energy levels. We therefore present a detailed review of the available experimental energy levels. The main results of this paper are $C_5$ and $C_6$ coefficients for all the Rydberg series where experimental energy levels are available. Tables of our complete results are provided in the supplementary data. In section \ref{Results} we present an overview of these tables, revealing significant differences between the species, and between different spin symmetries (singlet or triplet). Several F\"{o}rster resonances, namely accidental near-degeneracies in pair states of the two atoms causing a large value of the $C_6$ coefficient, are also identified. 

\section{Methods}
\label{Methods}
\subsection{Theory}
We consider two interacting divalent atoms, one in the state $n_1L_1S_1J_1$ and the other in the a state $n_2L_2S_2J_2$, separated by a distance $R$. Specifically, we consider states with high enough principal quantum numbers $n_1$ and $n_2$ that effects arising from configuration mixing can be neglected. Such effects, e.g., singlet-triplet mixing in the $5snd \; ^{1,3}D_2$ states of strontium \cite{Esherick1977a}, can be large in the vicinity of perturbers. However, in the species considered sharp deviations are only observed close to the perturber's energies. When far away from any perturber the Rydberg series varies monotonically, thus the effect of a perturber can be incorporated into the quantum defect. For strontium the perturbers are found much lower in the Rydberg series than the states considered here, however two of the Rydberg series of calcium and ytterbium are considerably perturbed over a wide range of states. A single active electron treatment is questionable for the calculating the $C_5$ and $C_6$ coefficients affected by these perturbations. The effect of these configuration interactions will be considered in future work. Here, we describe each atom by a simple model in which one of the two valence electrons is in an extended Rydberg orbital while the other is in a compact inner orbital. Denoting by ${\bf r}_a$ and ${\bf r}_b$ the position vectors of the valence electrons relative to the nucleus, we represent the state of each atom by the symmetrized wave function
\begin{align}
\sum_{M_{S},M_{L}} &\sum_{m_{l_{\rm i}}m_{l_{\rm o}}} \sum_{m_{s_{\rm i}}, m_{s_{\rm o}}} C_{L M_{L} S M_{S}}^{J M_{J}} C_{l_{\rm i} m_{l_{\rm i}} l_{\rm o} m_{l_{\rm o}}}^{L M_L}C_{s_{\rm i} m_{s_{\rm i}} s_{\rm o} m_{s_{\rm o}}}^{S M_S}\nonumber\\
\times &\left[\phi_{\rm i}({{\bf r}}_a)\chi_{m_{s_{\rm i}}}(a)\phi_{\rm o}({{\bf r}}_b)\chi_{m_{s_{\rm o}}}(b)\right.\nonumber\\
&\qquad + \left.(-1)^S \phi_{\rm i}({{\bf r}}_b)\chi_{m_{s_{\rm i}}}(b)\phi_{\rm o}({{\bf r}}_a)\chi_{m_{s_{\rm o}}}(a)\right].
\label{symmetrized}
\end{align}
The subscripts i and o refer to the inner and outer orbitals, $\chi_{m_{s_{}}}(a)$ and $\chi_{m_{s_{}}}(b)$ are spinors describing the spin states of electrons $a$ and $b$, $\phi_{\rm i}({\bf r})$ is the wave function of the inner electron, and $\phi_{\rm o}({\bf r})$ the wave function of the outer electron. We assume that the inner orbital is of $s$-symmetry; therefore $l_{\rm i}=m_{l_{\rm i}}=0$ and $l_{\rm o}=L$. 

The Hamiltonian for this system can be written as
\begin{equation}
\hat{H}= \hat{H}_0 + \hat{H}_{\mathrm{int}},
\label{hamiltonian}
\end{equation}
where $\hat{H}_0$ is the Hamiltonian of the pair of atoms at infinite separation and $\hat{H}_{\mathrm{int}}$ is the interaction energy between the two atoms. In terms of the position vectors of the valence electrons and the relative position ${\bf R}$ of the two nuclei
\begin{align}
\hat{H}_{\mathrm{int}} =&\sum_{\epsilon = a,b}\sum_{\epsilon' = a,b}\frac{1}{|{\bf r}_{\epsilon 1} - {\bf r}_{\epsilon' 2}-{\bf R}|} \nonumber\\
&- \sum_{\epsilon = a,b}\left(\frac{2}{|{\bf r}_{\epsilon 1} -{\bf R}|}   +\frac{2}{|{\bf r}_{\epsilon 2} +{\bf R}|}\right) + \frac{4}{R},
\label{hamiltonianint}
\end{align}
with the indexes 1 and 2 identifying the respective atoms, and $R=|{\bf R}|$. (Atomic units are used throughout this section). 

We treat the interaction Hamiltonian $\hat{H}_{\mathrm{int}}$ perturbatively.
For long range Rydberg-Rydberg interactions, the matrix elements of $\hat{H}_{\mathrm{int}}$ between the bound eigenstates of $\hat{H}_0$ are dominated by the contribution of the two outer electrons. We can therefore reduce the calculation of the dispersion coefficients to a problem in which each atom has only one active electron and the interaction Hamiltonian reduces to
\begin{equation}
\hat{H}_{\mathrm{int}}'=\frac{1}{|{\bf r}_{1} - {\bf r}_{2}-{\bf R}|}- \frac{1}{|{\bf r}_{1} -{\bf R}|} - \frac{1}{|{\bf r}_{2} -{\bf R}|} + \frac{1}{R},
\label{hamiltonianintprime}
\end{equation}
where ${\bf r}_1$ and ${\bf r}_2$ denote the position vectors of the two Rydberg electrons respective to the corresponding nuclei. (Although each atom has only one active electron, the quantum numbers of the system remain those pertaining to the original multi-electron problem.) This formulation neglects exchange interactions, which is valid as long as $R$ much exceeds the LeRoy radius \cite{Leroy1973},
\begin{equation}
R_{\rm LR} = 2 \left( \langle r_1^2 \rangle^{\frac{1}{2}} + \langle r_2^2 \rangle^{\frac{1}{2}}\right).
\label{leroyradius}
\end{equation}
A similar single active electron treatment has been previously shown to yield accurate Stark maps for two-electron Rydberg atoms \cite{Millen2011,Zhi2001}.

Expanding $\hat{H}_{\rm int}'$ in multipoles yields \cite{Dalgarno1966}
\begin{align}
\hat{H}_{\mathrm{int}}' &=\sum^{\infty}_{k_1,k_2 =1} \frac{(-1)^{k_2}} {R^{k_1 + k_2 +1}}\nonumber\\&\times \sqrt{\frac{ (4 \pi)^{3}(2 k_1 + 2 k_2)!}{(2k_1+1)!(2k
_2+1)! (2k_1 + 2k_2 + 1)}}\nonumber\\
&\times \sum_{p=-(k_1 + k_2)}^{k_1 + k_2} \sum_{p_1 = -k_1}^{k_1} \sum_{p_2 = -k_2}^{k_2}C_{k_1 p_1 , k_2 p_2}^{k_1 + k_2 , p} \nonumber\\
&\times r_1^{k_1} r_2^{k_2} Y_{k_1, p_1}(\hat{r}_1) Y_{k_2, p_2}(\hat{r}_2)Y_{k_1 + k_2, p}(\hat{R}),
\label{multipolarinteraction}
\end{align}
where $\hat{R}$ is the unit vector along the internuclear axis.
For infinite atomic separation, the eigenenergies of the Hamiltonian $\hat{H}$ coincide with those of $\hat{H}_0$, which are sums of energies of unperturbed atomic states. These asymptotic eigenenergies are thus degenerate in $M_1$ and $M_2$. (There is no degeneracy if $J_1=J_2=0$ since in this case the magnetic quantum numbers $M_1$ and $M_2$ can take only one value.) $\hat{H}_{\rm int}'$ mixes states of different $M_J$ values and as a result splits the degenerate asymptotic energy levels into a number of sublevels. Each of the latter differs from its $R \rightarrow \infty$ limit by a sublevel-specific, $R$-dependent shift $\Delta E$.
Treating $\hat{H}_{\mathrm{int}}'$ perturbatively and making use of the multipolar expansion (\ref{multipolarinteraction}) leads to an expression of these shifts in the form of a series of inverse powers of $R$,
\begin{equation}
\Delta E= \sum^{}_{N} \frac{C_N}{R^{N}}.
\label{cncoeffs}
\end{equation}
For the systems we are concerned with, this expansion is dominated at large interatomic separations by the term in $1/R^5$, when $C_5 \not= 0$, which arises to first order in $\hat{H}_{\rm int}'$ from the quadrupole-quadrupole interaction ($k_1=k_2=2$ in the multipolar expansion
of $\hat{H}_{\mathrm{int}}'$), and the term in $1/R^6$, which arises to second order from the dipole-dipole interaction ($k_1=k_2=1$). 

The $C_5$ and $C_6$ coefficients are the eigenvalues of the $(2J_1+1)(2J_2+1)$ by $(2J_1+1)(2J_2+1)$ matrices ${\sf C}_5(\hat{R})$ and ${\sf C}_6(\hat{R})$ formed by the $M_J$-dependent, $\hat{R}$-dependent coefficients
\begin{align}
c_5 &(M'_1M'_2,M_1M_2;\hat{R}) = \nonumber\\ 
&D_{22}({\alpha}M'_1M'_2,{\alpha}M_1M_2;\hat{R}) R_{22}(n_1n_1\alpha,n_2n_2\alpha)
\label{c5expression}
\end{align}
and
\begin{align}
c_6 &(M'_1M'_2,M_1M_2;\hat{R}) =- \sum \frac{1}{\Delta} \nonumber\\
 \times & D_{11}(\alpha M'_1M'_2,\alpha'' M''_1M''_2;\hat{R})R_{11}(n_1 n_2 \alpha , n''_1 n''_2 \alpha'')\nonumber\\
\times &D_{11}(\alpha'' M''_1M''_2, \alpha M_1M_2;\hat{R})R_{11}(n''_1 n''_2\alpha'' , n_1 n_2\alpha). 
\label{c6expression}
\end{align}
In these two equations, $\alpha$ denotes the sextuple of quantum numbers $L_1S_1J_1L_2S_2J_2$ and
\begin{align}
R_{k_1k_2} &(n_1 n_2 \alpha , n'_1 n'_2 \alpha') =\nonumber \\
 \int_0^{\infty} &dr_1  P_{n_1 L_1 S_1J_1}(r_1) r_1^{k_1} P_{n'_1 L'_1 S'_1 J'_1}(r_1)\nonumber \\
\times \int_0^{\infty} &dr_2  P_{n_2 L_2 S_2J_2}(r_2) r_2^{k_2} P_{n'_2 L'_2 S'_2 J'_2}(r_2).
\label{radialmatels}
\end{align}
The functions $D_{11}$ and $D_{22}$ are defined in the appendix.
Moreover, in equation (\ref{c6expression}) the sum runs over all the intermediate
$n_1''  n_2'' \alpha'' M_1''M_2''$ pair states dipole coupled both to the
$n_1 n_2 \alpha M_1 M_2$ state and to the
$n_1 n_2 \alpha M_1' M_2'$ state, and $\Delta$ denotes the F\"orster defect ($\Delta= E_1'' + E_2'' - E_1-E_2$). 

It follows from equation (\ref{multipolarinteraction}) that $\hat{H}_{\mathrm{int}}'$ only couples pair states of same value of $M_1+M_2$, and that the ensuing energy shifts do not depend on the overall sign of $M_1+M_2$, when the angle $\theta$ between the interatomic axis and the axis of quantization of the angular momenta (which we take to be the $z$-axis) is zero. The sublevels the asymptotic energy levels split into are thus characterized by $|\Omega|$, where $\Omega=M_1+M_2$. ($\Omega$ can be recognized as the magnetic quantum number associated with $\hat{J}_{{\rm t}z}$, the $z$-component of the total angular momentum operator ${\hat{\bf J}}_{\rm t}= {\hat{\bf J}}_1+{\hat{\bf J}}_2$.) For other orientations of the interatomic axis, $\hat{H}_{\mathrm{int}}'$ also couples pair states differing in $\Omega$. However, the choice of the quantization axis is arbitrary and thus the eigenenergies of the system do not depend on $\hat{R}$: changing the orientation of the interatomic axis changes the composition of the eigenstates of the Hamiltonian in terms of the unperturbed pair states but does not affect the energy shifts $\Delta E$. Therefore the dispersion coefficients as defined by equation (\ref{cncoeffs}) do not depend on $\hat{R}$ and can be obtained by diagonalizing the matrices ${\sf C}_5(\hat{R})$ and ${\sf C}_6(\hat{R})$ for any orientation of the interatomic axis. 

Due to the selection rule mentioned at the beginning of the last paragraph, these two matrices are block diagonal when this axis is in the $z$-direction, each block being formed by pair states with the same value of $\Omega$. It follows from the relationships between $\Omega$, $\hat{\bf J}_{\rm t}$, ${\hat{\bf J}}_1$ and ${\hat{\bf J}}_2$ that the linear size of each block, i.e., the number of values of $C_5$ or $C_6$ associated with
each value of $\Omega$, is $J_1+J_2-\max(|\Omega|,|J_1-J_2|)+1$. In particular, 
 the diagonal blocks with $\Omega= \pm (J_1+J_2)$ contain only one element. For these values of $\Omega$, the dispersion
coefficients are thus given directly by equations (\ref{c5expression}) and (\ref{c6expression}) as
\begin{eqnarray}
C_5&=&c_5(J_1J_2,J_1J_2;\hat{R}=\hat{z})\\
C_6&=&c_6(J_1J_2,J_1J_2;\hat{R}=\hat{z}).
\end{eqnarray}
 ($\Omega=\pm (J_1+J_2)$ implies that each atom is in a stretched state with $M_1=\pm J_1$ and $M_2=\pm J_2$.) 

\begin{table*}
\begin{ruledtabular}
\begin{tabular}{@{}lcccl}
Series & $\Omega$ & $K$ & Notes & Eigenstate\\
\hline
$^1S_0$, $^3P_0$ &  $0$ & $0$ & a,b & $|0,0\rangle$\\
\hline
$^1P_1$, $^3 P_1$, $^3 D_1$&  $0$ & $0$  & a,c & $\left( |1,-1\rangle + |-1, 1\rangle - |0,0\rangle\right)/\sqrt{3}$\\
 &   $0$ & $1$ & a,b& $\left( |1,-1\rangle - |-1,1\rangle\right)/\sqrt{2}$\\
 &   $0$ & $2$ & c&$\left(|1,-1\rangle +|-1,1\rangle \right)/\sqrt{6} + \sqrt{2/3}\,|0,0\rangle$\\
 &   $1$ & $1$ & a,b& $\left(|1,0\rangle - |0,1\rangle \right)/\sqrt{2}$\\
 &  $1$ & $2$ & b  &$\left(|1,0\rangle + |0,1\rangle \right)/\sqrt{2}$\\
 &  $2$ & $2$ & b & $|1,1\rangle$\\
\hline
$^1D_2$, $^3P_2$, $^3D_2$ &  $0$ & $0$ &  & $0.4320\left( |2,-2\rangle + |-2,2\rangle\right)
- 0.5593\left( |1,-1\rangle + |-1,1\rangle\right)\
+ 0.0331 \, |0,0\rangle$\\
&  $0$ & $1$&  & $0.3717 \left(|2,-2\rangle - |-2,2\rangle \right)
-0.6015 \left( |1,-1\rangle - |-1,1\rangle\right)$\\
&  $0$ & $2$ &  & $0.2064 \left( |2,-2\rangle + |-2,2\rangle \right)
+ 0.1316 \left( |1,-1\rangle + |-1,1\rangle\right)
- 0.9382\, |0,0\rangle$\\
&  $0$ & $3$ &  & $0.6015 \left( |2,-2\rangle - |-2,2\rangle \right)
+ 0.3717 \left(|1,-1\rangle - |-1,1\rangle\right)$\\
&  $0$ & $4$ &  & $0.5204\left(|2,-2\rangle + |-2,2\rangle \right)
+ 0.4121 \left(|1,-1\rangle + |-1,1\rangle \right)
+ 0.3445 \, |0,0\rangle$\\
&  $1$ & $1$ &   & $0.6971 \left( |2,-1\rangle - |-1,2\rangle \right)
+0.1184 \left( |0,1\rangle - |1,0\rangle\right)$\\
&  $1$ & $2$ &   & $0.6250 \left(|2,-1\rangle + |-1,2\rangle \right)
-0.3307 \left(|0,1\rangle + |1,0\rangle \right)$\\
&  $1$ & $3$ &   & $0.1184 \left(|2,-1\rangle - |-1,2\rangle \right)
+ 0.6971 \left(|1,0\rangle - |0,1\rangle \right)$\\
&  $1$ & $4$ &  & $0.3307 \left(|2,-1\rangle + |-1,2\rangle \right)
+ 0.6250 \left(|1,0\rangle + |0,1\rangle \right)$\\
&  $2$ & $2$ &   & $0.2810 \left(|2,0\rangle + |0,2\rangle \right)
- 0.9177 \, |1,1\rangle$\\
&  $2$ & $3$ & b  & $\left(|2,0\rangle - |0,2\rangle\right)/\sqrt{2}$\\
&  $2$ & $4$ &   & $0.6489 \left(|2,0\rangle + |0,2\rangle \right)
+ 0.3974 \, |1,1\rangle$\\
&  $3$ & $3$ &b  & $\left( |1,2\rangle -|2,1\rangle\right)/\sqrt{2}$\\
&  $3$ & $4$ &b   & $\left( |1,2\rangle +|2,1\rangle\right)/\sqrt{2}$\\
&  $4$ & $4$ &b & $|2,2\rangle$\\
\end{tabular}
\end{ruledtabular}
\caption{\label{symmetrytable}The eigenstates of the $\hat{H}^{(5)}$ Hamiltonian in terms of the $|M_1, M_2\rangle$ Zeeman substates of the pair states, for the case where the interatomic axis is aligned with the $z$-axis. It is assumed that $L_1=L_2$, $S_1=S_2$ and $J_1=J_2$. The eigenstates listed in this table are the same as those given in Table 1 of Ref.\ \cite{Singer2005}.
Note a: $C_5 = 0$ for this state. Note b: The state specified in the right-hand column is an eigenstate of $\hat{J}_{\rm t}^2$ and of $\hat{H}^{(6)}$ as well as of $\hat{H}^{(5)}$ for any $n$. Note c: The $K=0,\Omega=0$ and $K=2,\Omega=0$ eigenstates of $\hat{H}^{(5)}$ are also eigenstates of $\hat{J}_{\rm t}^2$ for $J=1$; however, the $K=0,\Omega=0$ and $K=2,\Omega=0$ eigenstates of $\hat{H}^{(6)}$ are not.}
\end{table*}

The calculation thus amounts to diagonalizing the Hamiltonians
\begin{eqnarray}
\hat{H}^{(5)}= \hat{H}_0 + \sum_{M'_1M'_2,M_1M_2} 
&c_5(M'_1 M'_2,M_1M_2;\hat{R})/R^5 \nonumber \\
&\times |M'_1M'_2\rangle\langle M_1M_2|
\end{eqnarray}
and
\begin{eqnarray}
\hat{H}^{(6)}= \hat{H}_0 + \sum_{M'_1M'_2,M_1M_2} &c_6(M'_1M'_2, M_1M_2;\hat{R})/R^6 \nonumber \\
&\times|M'_1M'_2\rangle\langle M_1M_2|
\end{eqnarray}
in the basis of the Zeeman substates $|M_1M_2\rangle$ of the pair state 
$n_1 n_2 \alpha$. The components of the eigenvectors of the matrices ${\sf C}_5(\hat{R})$ or ${\sf C}_6(\hat{R})$ are the coefficients of the expansion of the eigenvectors of $\hat{H}^{(5)}$ or $\hat{H}^{(6)}$ in this basis. For the first order quadrupole interaction, these coefficients are entirely determined by the angular factors $D_{22}({\alpha}M'_1M'_2,{\alpha}M_1M_2;\hat{R})$ and are independent of $n_1$ and $n_2$. They are 
given in table \ref{symmetrytable} for the cases of interest in this work. Because the $c_5$ functions depend on $S$ only through an overall factor, the eigenstates of $\hat{H}^{(5)}$ are the same for singlet and triplet states. They are identical to those obtained in Ref.\ \cite{Singer2005} for a spinless alkali atom.

That $\Omega$ is a good quantum number when $\hat{R}=\hat{z}$ originates from the fact that $\hat{J}_{{\rm t}z}$ commutes with $\hat{H}_{\rm int}'$ for this particular orientation of the interatomic axis. In contrast $\hat{J}_{\rm t}^2$ does normally not commute with $\hat{H}_{\rm int}'$. Nonetheless, as indicated in table \ref{symmetrytable}, some of the eigenstates of $\hat{H}^{(5)}$ for $\hat{R}=\hat{z}$ are also eigenstates of $\hat{J}_{\rm t}^2$. In view of this fact, we label the simultaneous eigenstates of $\hat{H}^{(5)}$ and $\hat{J}_{\rm t}^2$ by a number $K$ such that the corresponding eigenvalues of $\hat{J}_{\rm t}^2$ are $\hbar^2 K(K+1)$. For eigenstates of $\hat{H}^{(5)}$ that are not eigenstates of $\hat{J}_{\rm t}^2$, we assign the number $K$ to the eigenstate whose components in the Zeeman basis are closest to those of the eigenstate of $\hat{J}_{\rm t}^2$ with eigenvalue $\hbar^2 K(K+1)$ for the same $\Omega$. Doing so leads to the assignments indicated in table \ref{symmetrytable}. One may observe that the eigenstates are symmetric upon the interchange of the states of atom 1 and atom 2 for $K$ even and antisymmetric for $K$ odd. (The relevance of $\hat{J}_{\rm t}^2$ in this context had already been recognized in reference \cite{Walker2008}.)

We label the eigenstates of $\hat{H}^{(6)}$ in the same fashion. As noted in table \ref{symmetrytable}, several of these eigenstates are also eigenstates of $\hat{H}^{(5)}$ and of $\hat{J}_{\rm t}^2$. However, this is not the case in general. In particular, many of the eigenstates of $\hat{H}^{(6)}$ vary with $n_1$ and $n_2$, unlike the eigenstates of $\hat{H}^{(5)}$. The correspondence between even and odd values of $K$ and the symmetry under the interchange of the states of atoms 1 and 2 is nonetheless the same.

In general, the composition of a pair state $|M_1M_2\rangle$ in terms of energy eigenstates will therefore depend both on the orientation of the internuclear axis and on $R$, although the dependence on $R$ will be negligible when the expansion (\ref{cncoeffs}) is completely dominated either by the $1/R^6$ term or by the $1/R^5$ term. Due to the differences in the eigenvectors of $\hat{H}^{(5)}$ and of $\hat{H}^{(6)}$, the total energy shift at the interatomic distances where $|C_5/R^5| \approx |C_6/R^6|$ is best calculated by diagonalizing the Hamiltonian
\begin{align}
\hat{H}^{(5+6)}= \hat{H}_0 + \sum_{M'_1M'_2,M_1M_2} [c_5(&M'_1M'_2, M_1M_2;\hat{R})/R^5 \nonumber \\ +
c_6(M'_1M'_2, M_1M_2;\hat{R})/R^6&] \,
|M'_1M'_2\rangle\langle M_1M_2|.
\label{eq:H56}
\end{align}
 
Obtaining the dispersion coefficients thus largely reduces to a computation of radial matrix elements and of angular terms. We calculate the former using the analytical expressions derived in references \cite{Edmonds1979, Picart1978, Oumarou1988} in the framework of the Coulomb approximation. This approach yields accurate results for sufficiently high principal quantum numbers and does not require any other input than orbital angular momentum quantum numbers and binding energies. The latter are obtained from experimental data.

\begin{figure}[htbp]
\centering
\includegraphics{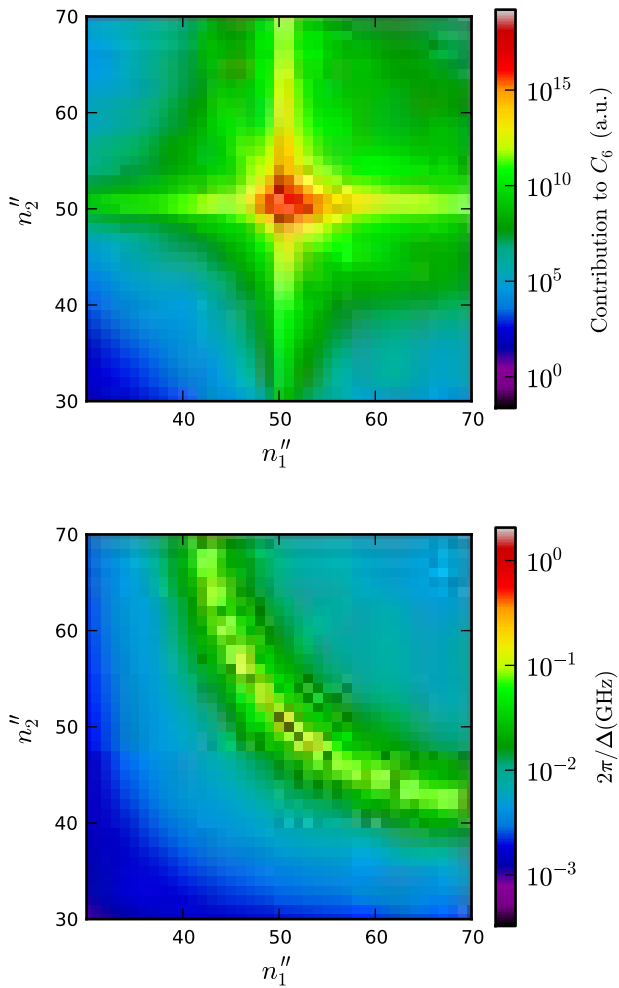}
\caption{(colour online) Contribution of individual intermediate $^1S_0$ pair states to the $C_6$ coefficient of the $5s50p \; ^1P_1$ state of strontium (top), and the inverse of the corresponding F\"orster defect (bottom). }
\label{convergence}
\end{figure}

The calculation of the $C_6$ coefficients also involves a summation over intermediate pair states. As is illustrated by figure \ref{convergence}, the sum is dominated by the pair states with the smallest F\"orster defect $\Delta$. Including in the sum the 15,000 (or thereabout) pair states with the smallest values of $\Delta$ was sufficient to ensure convergence of the $C_6$ coefficients to four significant figures. (The intermediate states included were restricted to principal quantum numbers in the range $10 \le n_1'', n_2'' \le 100$.) This rate of convergence is similar to that observed in the alkali metals \cite{Singer2005}.

Before closing this section, we briefly comment on the applicability of the perturbative approach at the interatomic separations relevant for cold Rydberg gases experiments (typically 2 to 10 $\mu$m). A comparison between a non-perturbative calculation of the energy shift due to the dipole-dipole interaction and the prediction of the leading-order perturbative calculation ($\Delta E = C_6/R^6$) is shown in figure 2. The non-perturbative shift was calculated by diagonalizing an Hamiltonian matrix of components
\begin{equation}
H_{pq} = \delta_{pq} {\cal E}_{p} + V_{pq},
\label{interactionmatrix}
\end{equation}
where the indexes $p$ and $q$ run over all the pair states included in the calculation, ${\cal E}_p$ is the asymptotic energy of the pair state $p$, and
\begin{align}
 V_{pq} = D_{11}&(\alpha_q M_{1q},M_{2q},\alpha_p M_{1p} M_{2p};\hat{R}=\hat{z}) \nonumber \\*
&\qquad \times R_{11} ( n_{1p} n_{2p}\alpha_p; n_{1q} n_{2q} \alpha_q)/R^{3}.
\label{dipoleinteraction}
\end{align}
For the state considered in figure 2, the perturbative $1/R^6$ shift matches the non-perturbative result very well for $R$ larger than about 1.5 $\mu$m but there are large differences at smaller separations. (The barely visible differences noticeable at larger values of $R$ originate from differences in the set of states taken into account: only 4,000 pair states were included in the non-perturbative calculation.) Perturbation theory breaks down when $\Delta E$ is comparable to or exceeds the F\"orster defect with the nearest pair state, $\Delta$, i.e., at a separation $R_{\rm np}$ such that $C_6/R_{\rm np}^6 \approx \Delta$. Hence $R_{\rm np}$ scales with the principal quantum number approximately like $n^{7/3}$: increasing $n$ from 50 to 100 increases $R_{\rm np}$ by about a factor 5, which may preclude the use of the corresponding $C_5$ and $C_6$ coefficients in systems where the typical interatomic spacing is a few microns. It is clear that nonperturbative effects become increasingly important at high $n$ and may be predominant at or near F\"orster resonances, where the $C_6$ coefficient is large and the F\"orster defect $\Delta$ is small. (The energy shift due to the quadrupole-quadrupole interaction, represented by a dashed red curve in figure 2, will be discussed in Section \ref{Results}.)

\begin{figure}[htbp]
\centering
\includegraphics{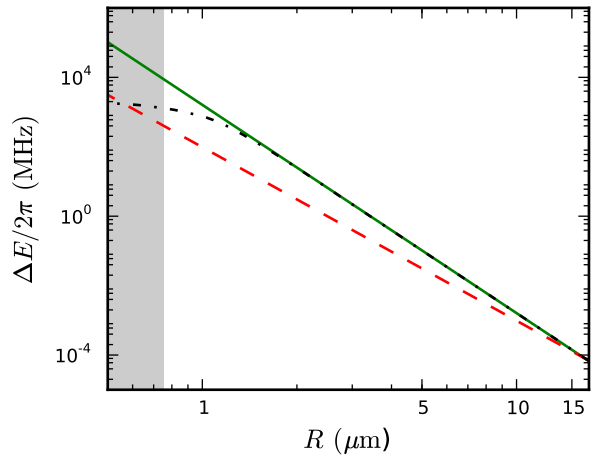}
\caption{(colour online) Comparison between the energy shift calculated non-perturbatively (black dash-dotted curve), the $C_6/R^6$ energy shift (green solid curve) and the $C_5/R^5$ energy shift (red dashed curve) for a pair of strontium atoms, both in the $5s50p \; {}^1 P_1$ state, at a distance $R$ from each other. (The results shown are for the $K=2,\Omega=0$ eigenstate, the only $\Omega=0$ pair state for which $C_5$ is non-zero). The shaded area represents the region where $R$ is smaller than the LeRoy radius.}
\label{nonpertcomparison}
\end{figure}

\subsection{Energy Levels}\label{energysection}
The calculation of the $C_5$ and, particularly, the $C_6$ coefficients requires an accurate knowledge
of the binding energies of all the relevant atomic states. However, the quality of the available experimental binding energies varies from series to series, as some are known only through laser spectroscopy measurements while others have been measured using microwave spectroscopy. We have used the microwave measurements where available, in view of their normally higher degree of accuracy. The random errors on the energy levels obtained by laser
spectroscopy are often large enough to affect the $C_6$ coefficients significantly. To reduce their impact, we fit the corresponding energy levels to the Rydberg-Ritz formula for the quantum defect \cite{Gallagher1994}. Thus, for a state of principal quantum number $n$, binding energy $E_{\rm b}$ and Rydberg constant $R_\mathrm{a}$, we express $E_{\rm b}$ in terms of the quantum defect,
$\delta = n- [R_\mathrm{a}/E_{\rm b}]^{1/2}$, and write 
\begin{equation}
\delta = \delta_0 + \frac{\delta_2}{(n-\delta_0)^2} + \frac{\delta_4}{(n-\delta_0)^4} + \dots
\label{rydbergritz}
\end{equation}
We set $\delta_k=0$ for $k> 4$ and obtain $\delta_0$, $\delta_2$ and
$\delta_4$ by least-square fitting to the data. The resulting values of these coefficients
are shown in table \ref{strontiumenergy} together with the ranges of principal quantum numbers used in the fit. The uncertainties quoted in the tables were obtained by considering the variation of the $\chi^2$ function about its minimum \cite{Hughes2010}. Correlations in these uncertainties,
given by the off-diagonal components of the covariance matrix, are negligibly small. We used the experimental energies directly for strongly perturbed series that could not be fitted in this way and the published quantum defects for the series measured by microwave spectroscopy.

In the case of strontium, no microwave measurements of energy levels are available but
laser measurements have been made for a number of series over wide ranges of values of $n$ \cite{Esherick1977a,Beigang1982a,Rubbmark1978,Beigang1982b,Armstrong1979,Esherick1977b,Beigang1983,Moore1952,Garton1968,Dai1995}. We used the energies of references \cite{Beigang1982a,Rubbmark1978,Beigang1982b,Armstrong1979,Esherick1977a} in view of their higher accuracy. The values of the $\delta_0$, $\delta_2$ and $\delta_4$ coefficients are given in Table \ref{strontiumenergy}. The Rydberg-Ritz formula fits the data well for all the series, with the exceptions of the $^3P_2$ and $^1S_0$ series for which substantial departures were found for high principal quantum numbers. (In the case of the $^1S_0$ states, the departure has been attributed to collisional shift with foreign gas \cite{Beigang1982a}.) For these two series, the fit was restricted to the values of $n$ over which the Rydberg-Ritz formula could match the data. The $^1D_2$ and $^3D_2$ series exhibit strong configuration mixing \cite{Esherick1977a}; nevertheless the energy levels are well described by the Rydberg-Ritz formula for $n\ge20$.

\begin{table*}
\begin{ruledtabular}
\begin{tabular}{llddrcr}
Atom & Series & \text{$\delta_0$}& \text{$\delta_2$} & \text{$\delta_4$} & Fitted Range & Ref.\\*
\hline
Sr & $5sns \; {}^1S_0$ &3.26896 (2) & -0.138 (7) & 0.9 (6)& $14 \le n \le 34$ &\cite{Beigang1982a}\\*
&$5snp \; {}^1P_1$ &2.7295 (7) & -4.67 (4) & -157 (2)&$10 \le n \le 29$ &\cite{Rubbmark1978}\\*
&$5snd \; {}^1D_2$ &2.3807 (2) & -39.41 (6) & -109 (2)\text{$\times 10^1$}& $20 \le n \le 50$ & \cite{Beigang1982a}\\*
&$5snf \; {}^1F_3$ &0.089 (1) & -2.0 (2) &3 (2)\text{$\times 10^1$} &$10 \le n \le 25$ &\cite{Rubbmark1978}\\*
&$5sns \; {}^3S_1$ &3.371 (2) & 0.5 (2) &-1 (2)\text{$\times 10^1$} &$13 \le n \le 45$ & \cite{Beigang1982b}\\*
&$5snp \; {}^3P_2$ &2.8719 (2) & 0.446 (5)&-1.9 (1) &$8 \le n \le 18$ & \cite{Armstrong1979}\\*
&$5snp \; {}^3P_1$ &2.8824 (2) & 0.407 (5)& -1.3 (1) &$8 \le n \le 22$ & \cite{Armstrong1979}\\*
&$5snp \; {}^3P_0$ &2.8866 (1) & 0.44 (1)&-1.9 (1) &$8 \le n \le 15$ & \cite{Armstrong1979}\\*
&$5snd \; {}^3D_3$ &2.63 (1) & -42.3 (3)&-18 (1) \text{$\times 10^3$} &$20 \le n \le 45$ & \cite{Beigang1982b}\\*
&$5snd \; {}^3D_2$ &2.636 (5) & -1 (2) &-9.8 (9)\text{$\times 10^3$} &$22 \le n \le 37$ & \cite{Esherick1977a}\\*
&$5snd \; {}^3D_1$ &2.658 (6) & 3 (2)&-8.8 (7)\text{$\times 10^3$} & $20 \le n \le 32$ & \cite{Beigang1982b}\\*
&$5snf \; {}^3F_4$ &0.120 (1) & -2.4 (2)&12 (2)\text{$\times 10^1$} &$10 \le n \le 24$ & \cite{Rubbmark1978}\\*
&$5snf \; {}^3F_3$ &0.120 (1) & -2.2 (2)&12 (2)\text{$\times 10^1$} &$10 \le n \le 24$ & \cite{Rubbmark1978}\\*
&$5snf \; {}^3F_2$ &0.120 (1) & -2.2 (2)&12 (2)\text{$\times 10^1$} &$10 \le n \le 24$ & \cite{Rubbmark1978}\\*
\\*
Ca & $4sns \; {}^1S_0$ & \multicolumn{1}{r}{\it 2.337930 (3)} & \multicolumn{1}{r}{\it -3.96 (10)} & &&\cite{Gentile1990}\\*
&$4snp \; {}^1P_1$ &\multicolumn{1}{r}{\it 1.885584 (3)} & \multicolumn{1}{r}{\it -0.114 (3)} &\multicolumn{1}{r}{\it -23.8 (25)} &&\cite{Gentile1990}\\*
&$4snd \; {}^1D_2$ &\multicolumn{4}{c}{{Highly perturbed series}} & \cite{Gentile1990}\\*
&$4snf \; {}^1F_3$ &\multicolumn{1}{r}{\it 0.09864 (9)} & \multicolumn{1}{r}{\it -1.29 (9)} &\multicolumn{1}{r}{\it 36} &&\cite{Miyabe2006}\\*
&$4sns \; {}^3S_1$ &\multicolumn{1}{r}{\it 2.440956 (3)} & \multicolumn{1}{r}{\it 0.350 (3)} & &&\cite{Gentile1990}\\*
&$4snp \; {}^3P_2$ & 1.9549 (8) & 2.5 (1) &-16 (1) \text{$\times 10^1$} & $12 \le n \le 60$& \cite{Armstrong1979}\\*
&$4snp \; {}^3P_1$ & \multicolumn{1}{r}{\it 1.964709 (3)} & \multicolumn{1}{r}{\it 0.228 (3)} & & & \cite{Gentile1990}\\*
&$4snd \; {}^3D_2$ & 0.8859 (5)\text{$^{\mathrm{a}}$} & \multicolumn{1}{r}{\it 0.13 (4)}\text{$^{\mathrm{a}}$} & & & \cite{Gentile1990}\\*
&$4snd \; {}^3D_1$ & 0.8833 (5)\text{$^{\mathrm{b}}$} & \multicolumn{1}{r}{\it -0.02 (4)}\text{$^{\mathrm{b}}$} & & & \cite{Gentile1990}\\*
\\*
Yb &$4f^{14} 6sns \; {}^1S_0$ & \multicolumn{1}{r}{\it 4.27914 (4)} & \multicolumn{1}{r}{\it -7.06 (6)} & \multicolumn{1}{r}{\it 565 (25)} & &\cite{Maeda1992}\\*
&$4f^{14} 6snp \; {}^1P_1$ &\multicolumn{1}{r}{\it 3.95433 (5)} & \multicolumn{1}{r}{\it -12.33 (6)} & \multicolumn{1}{r}{\it 1729 (27)}& & \cite{Maeda1992}\\*
&$4f^{14} 6snd \; {}^1D_2$ & \multicolumn{1}{r}{\it 2.71363 (4)} & \multicolumn{1}{r}{-2.01 (4)} &  & &\cite{Maeda1992}\\*
&$4f^{14} 6snf \; {}^1F_3$ & \multicolumn{4}{c}{{Highly perturbed series}} &  \cite{Aymar1984}\\*
\end{tabular}
\end{ruledtabular}
\caption{\label{strontiumenergy}The Rydberg-Ritz parameters for strontium, calcium and ytterbium used in the calculation of the $C_5$ and $C_6$ coefficients. Uncertainties in the last digits are given in brackets. The parameters printed in italic are quoted from the sources given in the last column of the table. The others were obtained in this work. For those, the last column gives the reference to the sources of the spectroscopic data used in the calculation and the fifth column the range of principal quantum numbers included in the fit. The Rydberg constants for strontium, calcium and ytterbium are $R_{\mathrm{Sr}}= 109\,736.627 \mathrm{cm}^{-1}$ \cite{Beigang1982a}, $R_{\mathrm{Ca}}= 109\,735.81 \mathrm{cm}^{-1}$ \cite{Gentile1990} and $R_{\mathrm{Yb}}= 109\,736.96 \mathrm{cm}^{-1}$ \cite{Baig1992} respectively.
\newline ${}^\mathrm{a}$Due to a perturber, a term $9.08 (9) \times 10^{-4} \left[(n-\delta_0)^{-2} - 0.01676700 \right]^{-1}$ must be added to equation (\ref{rydbergritz}) for this series.
\newline ${}^\mathrm{b}$Due to a perturber, a term $8.51 (9) \times 10^{-4} \left[(n-\delta_0)^{-2} - 0.01685410 \right]^{-1}$ must be added to equation (\ref{rydbergritz}) for this series. }
\end{table*}

Microwave spectroscopy measurements have provided very precise energy levels for calcium \cite{Gentile1990,Miyabe2006} and ytterbium \cite{Maeda1992}. We have supplemented these results with laser spectrosopy measurements of the $^3P_2$ series of calcium \cite{Armstrong1979} and of the
$^1F_3$ series of ytterbium \cite{Aymar1984}. Altogether, though, fewer series have been measured for these elements than for strontium (table \ref{strontiumenergy}). Moreover, the $^1D_2$ series of calcium and the $^1F_3$ series of ytterbium, which are highly perturbed, cannot be fitted to the Rydberg-Ritz formula; as a consequence, we could not extrapolated the measured energies to other values of $n$. These limitations reduced the number of states for which we could calculate the $C_6$ coefficient.

\subsection{Uncertainties}Errors in the binding energies of the relevant states dominate the uncertainty on the dispersion coefficients for most series. The fit of the experimental quantum defects to the Rydberg-Ritz formula reduces this uncertainty in smoothing out the random scatter in the data. We estimated the uncertainty on the values of the $C_5$ and $C_6$ coefficients arising from the errors on the values of $\delta_0$, $\delta_2$ and $\delta_4$ by varying each of these parameters one by one and adding the resulting differences in quadrature \cite{Hughes2010}. (This procedure is likely to overestimate the total error for the series measured by microwave spectroscopy \cite{Gentile1990}.) The error introduced by extrapolating the Rydberg-Ritz formula to outside the range of principal quantum numbers used in the fits could not be ascertained.
 
Another source of error is the use of the Coulomb approximation to calculate the radial matrix elements.We have compared the Coulomb approximation for the $^1S_0$ series of strontium to a calculation of the $C_6$ coefficients using a model potential of the form
\begin{equation}
V_{\mathrm{model}}= -\frac{1}{r} \left[ 1 + (Z-1)e^{-\alpha r} + Br e^{-\beta r} \right].
\label{modelpotential}
\end{equation}
Here, $Z$ is the atomic number and $B$, $\alpha$ and $\beta$ are parameters obtained by least-squares fitting of the energy levels to the experimental data. The radial matrix elements were found to differ typically by about 0.1 a.u. between the two calculations, which translates to differences in the values of the $C_6$ coefficients of about 0.5\% for $n\approx 20$ and less for larger values of $n$. These differences are smaller than the uncertainty originating from the error in the energies for this series. However, it may be that the error introduced by the Coulomb approximation dominates the total error where all the relevant energies are known accurately from microwave measurements. 

\section{Results and Discussion}\label{Results}

Tables of the $C_5$ and $C_6$ coefficients for the series listed in table \ref{seriesindex} are provided in the on-line supplementary data accompanying this paper and form the main results of this work. We only consider pair states where both atoms are in the same Rydberg state --- thus $n_1=n_2=n$, $L_1=L_2=L$, $S_1=S_2=L$ and $J_1=J_2=J$. Tables of the $c_5(M'_1M'_2,M_1M_2;\hat{R})$ and the $c_6(M'_1M'_2,M_1M_2;\hat{R})$ functions are also provided in the Supplementary Information for $\hat{R}=\hat{z}$, i.e., for the case where the internuclear axis is aligned with the axis of quantization of the angular momenta. These results can be used, e.g., to obtain the eigenvectors of $\hat{H}^{(5)}$ and $\hat{H}^{(6)}$. Simple polynomial fits to selected $C_6$ coefficients are given in table \ref{c6tablesr}.

An overview of these results is presented in the next sections. We first examine the form of the long-range interaction, then consider the $C_5$ and $C_6$ coefficients in more detail.

\begin{table*}
\begin{ruledtabular}
\begin{tabular}{lcc}
Atom & Available $C_6$ and $c_6$ coefficients & Available $C_5$ and $c_5$ coefficients \\
\hline
Strontium & $^1S_0$ $^3S_1$ $^1P_1$ $^3P_{0,1,2}$ $^1D_2$ $^3D_{1,2,3}$ &  $^1P_1$ $^3P_{1,2}$ $^1D_2$ $^3D_{1,2,3}$\\
Calcium & $^1S_0$ ($^1P_1$) $^3P_{1}$ ($^1D_2$) &$^1P_1$ $^3P_{1,2}$ ($^1D_2$) $^3D_{1,2}$\\
Ytterbium & $^1S_0$ $^1P_1$ ($^1D_2$) & $^1P_1$ $^1D_2$\\
\end{tabular}
\end{ruledtabular}
\caption{\label{seriesindex} Index of the coefficients tabulated in the supplementary data. Results are normally given for $30 \le n \le 70$. However, only estimates for a reduced range of principal quantum numbers are given for the series indicated between brackets, due to a lack of spectroscopic data and the questionability of a single active electron treatment for this calculation.}
\end{table*}

\begin{table*}
\begin{ruledtabular}
\begin{tabular}{llcdddd}
Atom &Series & $|M_J|$ & $a$& $b$& $c$ & Fractional Error\\*
\hline
Sr&$5sns \; ^1S_0$&0&3.2\text{$\times 10^{-3}$}&-0.51&3.6&0.02\\*
&$5sns \; ^3S_1$&1&-2.387\text{$\times 10^{-3}$}&1.211&-21.18&0.01\\*
&$5snp \; ^1P_1$&1&-1.24\text{$\times 10^{-4}$}&0.0349&1.03&0.002\\* 
&$5snd \; ^1D_2$&2&-1.65\text{$\times 10^{-3}$}&0.365&-7.05&0.02\\* 
\hline
Ca&$4sns \; ^1S_0$&0&-1.793\text{$\times 10^{-3}$}&0.3190&-1.338&0.0001\\* 
\hline
Yb&$4f^{14}6sns \; ^1S_0$&0&-9.84\text{$\times 10^{-5}$}&0.0234&-0.421&0.003\\* 
&$4f^{14}6snp \; ^1P_1$&1&-7.74\text{$\times 10^{-4}$}&0.167&-2.73&0.0009\\* 
\end{tabular}
\end{ruledtabular}
\caption{\label{c6tablesr}Polynomial fits to the $C_6$ coefficients for stretched states of strontium, calcium and ytterbium ($M_1=M_2=M_J=\pm J$ with respect to the internuclear axis). $C_6 = n^{11} (a n^2 + b n + c)$ and the coefficients $a$, $b$ and $c$ are expressed in a.u. ($C_6\;\mathrm{(GHz\,\mu m^{6})} = 1.4448\times10^{-19}C_6\;\mathrm{(a.u.)}$.) These fits are valid for $30 \le n \le 70$. The fractional error quoted is the uncertainty in $C_6$ for $n=50$ due to the uncertainty in the energy levels. For all the series considered in the table this fractional error exceeds the deviation of the fitting polynomial from the calculated $C_6$ coefficients.}
\end{table*}

\subsection{Long-range interactions}

\begin{figure}[htbp]
\centering
\includegraphics{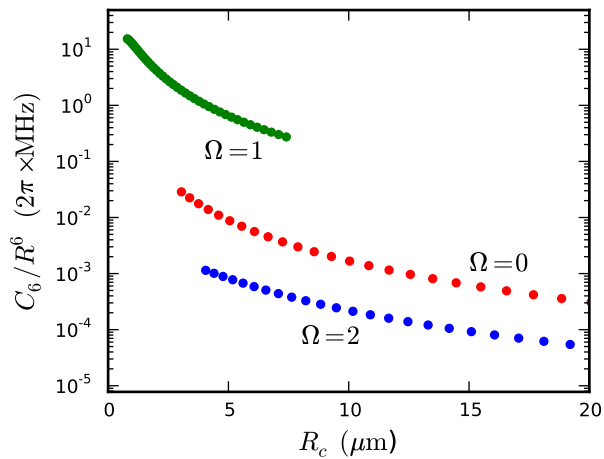}
\caption{(colour online) The energy shift $C_6/R^6$ at the critical radius $R_{\rm c}$ where $|C_5/R^5|=|C_6/R^6|$, for $5snp \; {}^1 P_1$ states of strontium. The principal quantum number, $n$, increases from $30$ to $70$ and from left to right in this figure. The total energy shift must be calculated by diagonalizing the Hamiltonian defined by equation
(\ref{eq:H56})
where $|C_5/R^5| \approx |C_6/R^6|$. The $C_5$ coefficient is negative and the $C_6$ coefficient is positive for $\Omega=1$, which means that for this symmetry the total energy shift vanishes at $R\approx R_{\rm c}$ and the long-range interaction changes from repulsive below $R_{\rm c}$ to attractive beyond $R_{\rm c}$.}
\label{energydistance}
\end{figure}

The relative strength of the first-order quadrupole-quadrupole interaction ($C_5$) and the second-order dipole-dipole interaction ($C_6$) is illustrated in figure \ref{nonpertcomparison}. As is well known, the $C_5/R^5$ term dominates the energy shift at large separation $R$ (for the symmetries for which $C_5\not= 0$), while the $C_6/R^6$ term is significant below a critical radius $R_{\rm c}=C_6/C_5$. How large $R_{\rm c}$ is depends on the symmetry of the eigenstate and on $n$, see figure \ref{energydistance} ($R_{\rm c}$ is roughly proportional to $n^3$). For most states, the $C_6/R^6$ term dominates the energy shift up to interatomic distances at which this shift is too small to be relevant for experiments. However for certain symmetries, such as the $\Omega=1$ state represented in the figure, the second order dipole-dipole shift is unusually small due to a vanishing angular factor \cite{Walker2008}. In this case, the quadrupole interaction may be of significant importance when considering a dipole blockade. 

Given that the quadrupole interaction can normally be neglected at the atomic densities involved in cold Rydberg gases experiments, we consider the $C_5$ coefficients only briefly in the following, before discussing the $C_6$ coefficients in greater detail.

\subsection{$C_5$ coefficients}
As they depend on the energy levels only through radial matrix elements, the $C_5$ coefficients exhibit less structure than the $C_6$ coefficients, which also depend on the inverse of the F\"orster defect, $1/\Delta$. Due to selection rules on the orbital angular momentum, the first order quadrupole interaction vanishes for many of the series. In particular, $C_5\equiv 0$ for $J = 0$.

In general, the $C_5$ coefficient scales with $n$ like $n^8$ \cite{Singer2005}. Apart for this scaling, these coefficients have a similar magnitude for most states. However, their sign varies from symmetry to symmetry. For the $J=1$ states, the $C_5$ coefficient is non-zero only for the $K=2$ eigenstates. The corresponding energy shifts are generally one order of magnitude smaller for the $K=2,\Omega=\pm 2$ eigenstates than for the $K=2,\Omega=\pm 1$ and $K=2,\Omega=0$ eigenstates, which have $C_5$ coefficients closer in magnitude but opposite in sign.

A large number of eigenstates have a non-vanishing quadrupole interaction for $J=2$. The corresponding values of $C_5$ tend to arrange themselves evenly about zero, without marked differences between strontium, calcium and ytterbium.

In general, most triplet states were found to have weaker quadrupole interactions than the singlet states for the same value of $n$, due to a smaller value of $D_{22}$.

\subsection{$C_6$ coefficients}
 
\begin{figure}[htb]
\centering
\includegraphics{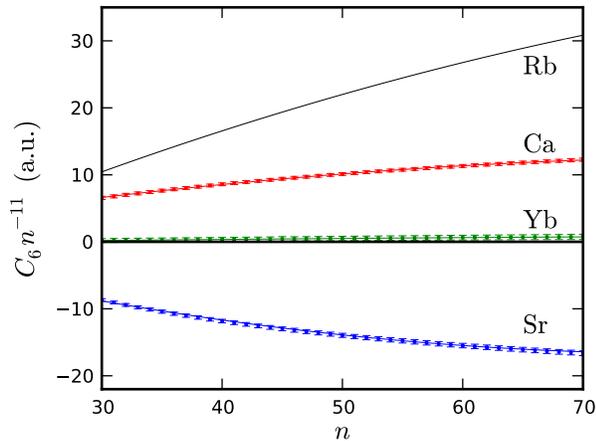}
\caption{(colour online) The scaled $C_6$ coefficient for the $^{2} S_{1/2}$ series of rubidium \cite{Singer2005} and for the $^1 S_0$ series of strontium, calcium and ytterbium. For the latter three atoms, the solid line represents the polynomial fit of Table \ref{c6tablesr}. The error bars are multiplied by 200 for calcium and ytterbium.}
\label{1s0series}
\end{figure}

First of all, we consider the Rydberg series with orbital angular momentum $L=0$. In alkali metals, the $S$ states are widely used in experiments as the interaction is repulsive, which reduces the effect of ionizing interactions, and is nearly independent of $\theta$. In contrast to the alkali metals, there are two $L=0$ Rydberg series in two-electron atoms. The $^1S_0$ states of the bosonic isotopes are particularly appealing for experiments as they have no magnetic sublevels. The $C_6$ coefficients for the $^1S_0$ series of strontium, calcium, ytterbium and rubidium are compared in figure~\ref{1s0series}. The three divalent atoms exhibit dramatically different behaviour. For Sr, the interaction is attractive\footnote{Anomalous behaviour in the Sr $5sns \; ^1S_0$ energy levels \cite{Beigang1982a} for $n>36$ (attributed to collisional shifts) leads to a discrepancy between the results in Table \ref{c6tablesr} and previous work \cite{Mukherjee2011}. Experimental values were used in \cite{Mukherjee2011}, while we fit the energy levels at $n<34$ where these effects are absent. The interaction is attractive in both cases.}
 while for Ca it is repulsive (although weaker than in Rb). The interaction is also repulsive in the case of Yb, however the scaled $C_6$ coefficients are over an order of magnitude smaller than for the other two species. These differences entirely arise from differences in the energy level spacing between these atoms.

The isotropically attractive interaction for strontium may have uses in many-body entangled states \cite{Mukherjee2011} and in non-linear self-focussing schemes \cite{Sevincli2011}. This interesting feature is not unique to this atom. We have estimated the $C_6$ coefficient for the $^1S_0$ series of magnesium, mercury and zinc using the quantum defects of references \cite{Rafiq2007,Zia2004,Nadeem2006,Kompitsas1994}, and found the interaction to be repulsive for magnesium and mercury but attractive for zinc. 

Also of note is the weak interaction in the ytterbium $^1 S_0$ states, which illustrates the importance of performing detailed calculations for each series. Such a weak interaction would be detrimental to any experiment wishing to exploit the $^1 S_0$ series of ytterbium to produce a Rydberg blockade.

\begin{figure}[htbp]
\centering
\includegraphics{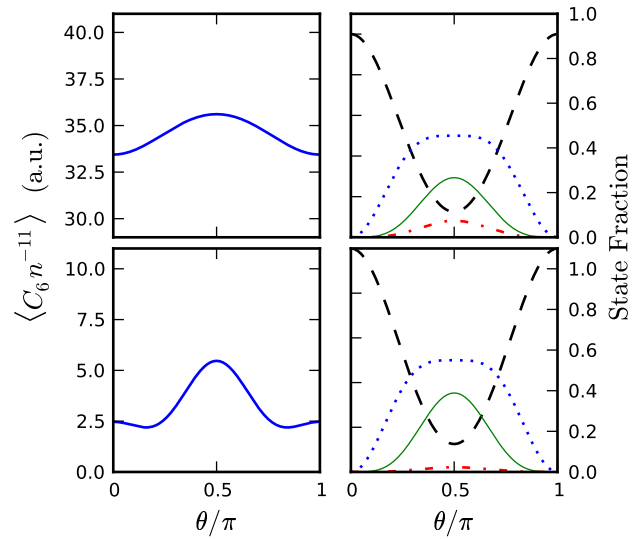}
\caption{(colour online) Left: The expectation value of the scaled $C_6$ coefficient for the $n=50$ stretched states of the
$^3S_1$ (top) and $^1P_1$ (bottom) series of Sr, $|1,1\rangle$, as a function of the angle $\theta$ between the direction of quantization of the angular momenta and the internuclear axis. Right: 
The composition in energy eigenstates of these two stretched states. Red dash-dotted curves: $K=0, \Omega=0$. Green solid curves: $K=2, \Omega=0$. Blue dotted curves: sum of $K=2, \Omega=1$ and $K=2,\Omega=-1$. Black dashed curves: sum of $K=2, \Omega=2$ and $K=2, \Omega=2$. The $K=1 , \Omega=0$ and $K=1,\Omega=\pm 1$ eigenstates are orthogonal to $|1,1\rangle$ as they have opposite symmetry under the interchange of the states of atoms 1 and 2.}
\label{angularstatefraction}
\end{figure}

Whereas the dipole-dipole interaction is attractive in the $^1 S_0$ series of strontium, it is repulsive in the $^3 S_1$ series. For these states, $\hat{H}^{(6)}$ has six different eigenenergies for each $n$, three of which are doubly degenerate. (Different eigenstates have different energy shifts because of the fine structure of the intermediate $^3 P$ states.) The resulting values of the scaled $C_6$ coefficient, $C_6n^{-11}$, range from 33.2 a.u.\ to 36.8 a.u.\ for $n=50$. The strength of the interaction varies thus little between these eigenstates and is always larger than for the $^2 S_{1/2}$ states of rubidium of same principal quantum number. These differences between eigenstates make the van der Waals interaction slightly anisotropic in the $^3 S_1$ series, in that controlled excitation to a particular $M_J$ state will generally excite a superposition of energy eigenstates whose composition will depend on the orientation of the internuclear axis of each pair of atoms in the cloud. Suppose, for example, that one prepares the two atoms of a pair in the same stretched state ($M_1=M_2=1$). This state is an eigenstate of $\hat{H}^{(6)}$ only when the internuclear axis is aligned with the axis of quantization of the angular momenta. Otherwise, its composition in terms of eigenstates of $\hat{H}^{(6)}$ varies with the angle $\theta$ between the two axes (figure \ref{angularstatefraction}). However, as seen from the figure, the average $C_6$ coefficient varies by less than 10\% between $\theta=0$ and $\theta=\pi/2$. This means that there are two nearly-isotropic $S$ series in strontium with $C_6$ coefficients of opposite signs, and therefore that the interaction can be tailored to experimental requirements by modifying the excitation scheme for the same atom.

The difference in $C_6$ coefficients between the different eigenstates of $\hat{H}^{(6)}$ is more significant in the $^1 P_1$ series than in the $^3 S_1$ series, leading to a larger relative variation of the expectation value of the energy on the state $|M_1=1,M_2=1\rangle$ (bottom row of figure \ref{angularstatefraction}). The composition of $|M_1=1,M_2=1\rangle$ in terms of the eigenstates of $\hat{H}^{(6)}$ is very similar for the two series --- it is actually identical for the $\Omega=\pm 1$ and $\Omega=\pm 2$ eigenstates; however, in the $^3 S_1$ states it is governed by the orientation of the spin of the four valence electrons and in the $^1 P_1$ states by the orientation of the two Rydberg $p$-orbitals.

\begin{figure}[htb]
\centering
\includegraphics{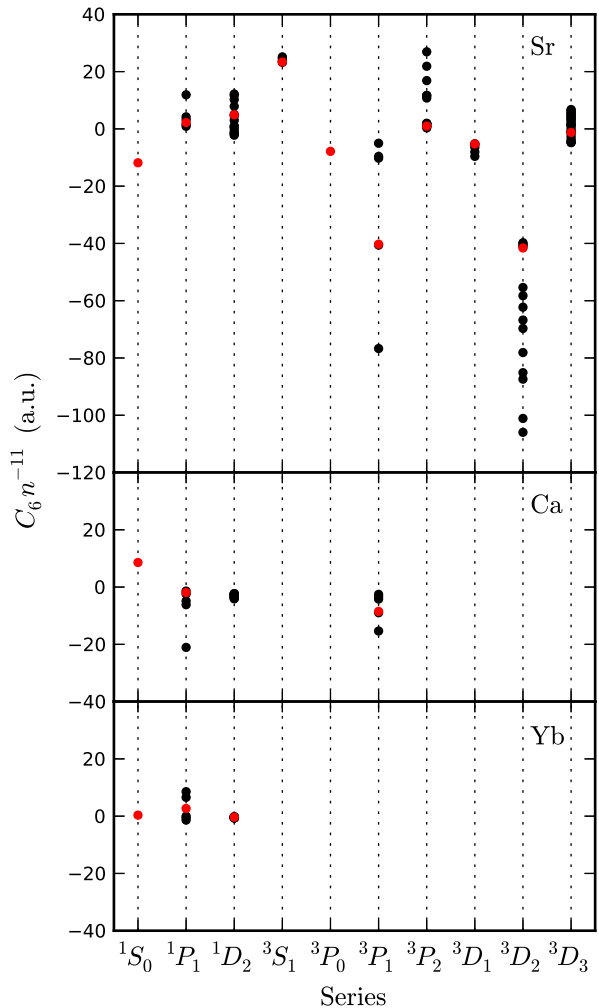}
\caption{(colour online) The scaled $C_6$ coefficient at $n=40$ for all the series and energy eigenstates considered in this work. The red markers indicate the value of $C_6$ in the stretched eigenstates ($|\Omega| = 2 J$). 
($C_6\;\mathrm{(GHz\,\mu m^{6})} = 1.4448\times10^{-19}C_6\;\mathrm{(a.u.)}$.)
}
\label{otherseries}
\end{figure}

\begin{figure}[htb]
\centering
\includegraphics{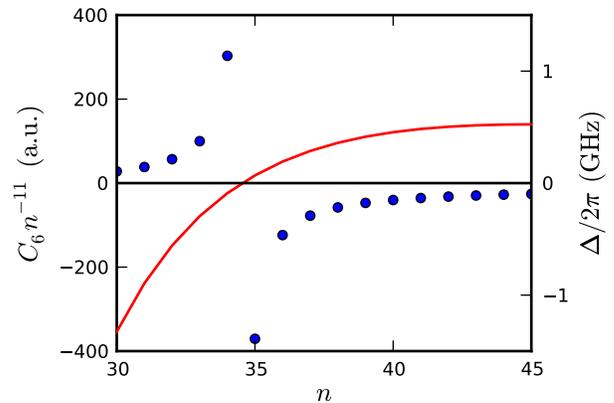}
\caption{(colour online) The $K=2,|\Omega|=2$ scaled $C_6$ coefficient for the Sr $5snp \; {}^3P_1$ series (blue circles), and the F\"orster defect for the $2\times 5snp\, {}^3P_1 \rightarrow 5sns\, {}^3S_1 + 5s(n+1)s\, {}^3S_1$ channel (red solid line).
($C_6\;\mathrm{(GHz\,\mu m^{6})} = 1.4448\times10^{-19}C_6\;\mathrm{(a.u.)}$.)}
\label{3p2seriessr}
\end{figure}

Figure \ref{otherseries} gives a snapshot of the strength and sign of the van der Waals interaction at $n=40$. In the cases of the $^1 D_2$ and $^3 D_3$ series of Sr and of the $^1 P_1$ series of Yb, the $C_6$ coefficient is positive for some of the eigenstates and negative for the others, and for certain values of $K$ and $\Omega$ its sign depends on $n$. These sign changes occur through the $C_6$ coefficient smoothly passing through zero as $n$ varies. (The $C_6$ coefficients are never exactly zero in our calculations, contrary to those discussed in Ref.\ \cite{Walker2008}, as we take into account all the intermediate angular channels.) The sign of $C_6$ also changes with $n$ in the $^3 P_1$ and $^3 D_2$ series of Sr, but in these two cases the change is abrupt and occurs at a F\"orster resonance, namely when the F\"orster defect of the dominant channel for the series changes sign and almost vanishes (figure \ref{3p2seriessr}). The corresponding near degeneracies are between the
$5s35p \, ^3P_1 + 5s35p \, ^3P_1$ and $5s35s \, ^3S_1 + 5s36s \, ^3S_1$ pair states ($\Delta/2\pi = 68$ MHz) and between the
$5s37d \, ^3D_2 + 5s37d \, ^3D_2$ and $5s34f \, ^3F_3 + 5s35f \, ^3F_3$ pair states ($\Delta/2\pi = 3$ MHz), respectively. These two resonances also give rise to abnormally large values of the $C_6$ coefficient in the $^3P_1$ and $^3D_2$ series of Sr, as can be noticed in figure \ref{otherseries}. We have not found F\"orster resonances in Ca or Yb, or in any other series of Sr for principal quantum numbers in the range $30 \leq n \leq 70$.

\section{Conclusions}
We have presented perturbative calculations of the long-range interaction between calcium, strontium and ytterbium Rydberg atoms. Extensive tables of the $C_5$ (first order quadrupole-quadrupole) and $C_6$ (second order dipole-dipole) coefficients are provided in the supplementary data. These calculations are based on the Coulomb approximation, and accurate binding energies are required as an input. The experimental energy level data currently available is summarized by the quantum defects listed in table \ref{strontiumenergy}.

The $1/R^6$ interaction is generally dominant in the range of interatomic separation important for experiments, although in channels where this interaction is unusually weak the quadrupole-quadrupole $1/R^5$ interaction may become significant. Comparing the $C_6$ coefficients for the same series revealed significant variations between the species, illustrating the fact that small differences in energy level spacings may have dramatic effects on Rydberg interactions. For example, the $1/R^6$ interaction for the $^1S_0$ series was found to be attractive for strontium, repulsive for calcium and nearly zero for ytterbium. The sign of the interaction can also change between different spin symmetries, as, e.g., in the $^1S_0$ and $^3S_1$ series of strontium. Significant variation was also observed for different symmetries within the same atomic species, in particular in the presence of a F\"orster resonance. Such resonances are less common in two-electron atoms than in alkali metals; only two instances were found in this work, both in triplet states of strontium.

\begin{acknowledgments}
The authors wish to acknowledge H James for her work on F\"orster resonances in strontium, calcium and ytterbium. They would also like to thank I G Hughes for helpful discussions and J Millen for preliminary work on strontium interactions. Financial support was provided by EPSRC grant EP/D070287/1.
\end{acknowledgments}

\appendix
\section{Angular Factors}
\label{angmatels}
Obtaining the $C_5$ and $C_6$ coefficients involves the calculation of angular matrix elements of
the interaction Hamiltonian $\hat{H}_{\rm int}$ between unperturbed states of the form given by
equation (\ref{symmetrized}). We collect the angular integrals together and define a coefficient $D_{k_{1}k_{2}}(\alpha' M'_1M'_2,\alpha M_1M_2;\hat{R})$ depending on the quantum numbers of the states, on the order of the multipole transition considered, and on the orientation of the internuclear axis. Here we present a derivation of these angular coefficients.

Recall that we assume that the inner electron is in a $s$-orbital. Hence, $l_{\rm i} = m_{l_{\rm i}}=0$, $L=l_{\rm o}$, $M_L= m_{l_{\rm o}}$ and $C_{l_{\rm i} m_{l_{\rm i}} l_{\rm o} m_{l_{\rm o}}}^{L M_L}=1$ and the angular part of the calculation reduces to that of the matrix elements $\langle  L' S' J' M_J' | Y_{k,p} |  L S J M_J \rangle$. However,
\begin{align}
 \langle  L' &S' J' M_J' | Y_{k,p} |  L S J M_J \rangle = \nonumber\\
&\sum_{m_{l_{\rm o}}, m_{l'_{\rm o}}} \sum_{M_S} \sum_{m_{s_{\rm i}}, m_{s_{\rm o}}} \delta_{m_{s_{\rm i}}, m_{s'_{\rm i}}} \delta_{m_{s_{\rm o}}, m_{s'_{\rm o}}} \delta_{M_{S}, M'_{S}}\nonumber\\
&\quad \times (C_{s_{\rm i} m_{s_{\rm i}} s_{\rm o} m_{s_{\rm o}}}^{S M_s})^2 C_{l_{\rm o} m_{l_{\rm o}} S M_{S}}^{J M_J} C_{l'_{\rm o} m'_{l_{\rm o}} S' M'_{S}}^{J' M'_J}\nonumber\\
&\qquad \times \langle  l_{\rm o}' m'_{l'_{\rm o}} | Y_{k,p} |  l_{\rm o} m_{l_{\rm o}} \rangle \nonumber\\
&= \sum_{m_{l_{\rm o}}, m_{l'_{\rm o}}}\sum_{M_S} C_{l_{\rm o} m_{l_{\rm o}} S M_{S}}^{J M_J} C_{l'_{\rm o} m'_{l_{\rm o}} S M_{S}}^{J' M'_J}\nonumber\\
&\qquad  \times\langle  l'_{\rm o} m'_{l'_{\rm o}} | Y_{k,p} |  l_{\rm o} m_{l_{\rm o}} \rangle.
\label{matrixelement1}
\end{align}
Since $\hat{H}_{\rm int}'$ does not couple singlet to triplet states, $S_1=S_1'$ and $S_2=S_2'$.
Using standard summation rules and evaluating the angular components of the matrix elements \cite{Messiah1961} yields the familiar result
\begin{align}
\langle  L' S' J' M'_J &| Y_{k,p} |  L S J M_J \rangle\nonumber\\
& = (-1)^{l_{\rm o} + l'_{\rm o} +1} \sqrt{(2l_{\rm o} + 1) (2J+1) (2k+1) \over 4\pi}\nonumber\\
&\qquad \times  C_{l_{\rm o} 0 k 0}^{l'_{\rm o} 0} C_{J M_J k p}^{J' M'_J}
\left\lbrace\begin{array}{ccc}
J & k & J'\\
l'_{\rm o} & S & l_{\rm o}
\end{array}\right\rbrace.
\label{matrixelement2}
\end{align}
Making use of these results and of equation (\ref{multipolarinteraction}), we express
the matrix elements of the Hamiltonian $\hat{H}_{\rm int}'$ as sums of radial terms multiplied by the coefficients
\begin{align}
 D_{k_{1}k_{2}}&(\alpha'M'_1M'_2,\alpha M_1M_2;\hat{R}) = (-1)^{k_2} \nonumber\\
&\times \sqrt{\frac{4\pi (2k_1 + 2k_2)!(2L_1 +1)(2L_2 +1) }{(2k_1)!(2k_2)!(2k_1 + 2k_2 +1)}}\nonumber\\
  &\times \sqrt{(2J_1 +1) (2J_2 +1)}\,C_{L_1 0,k_1 0}^{L'_1 0} C_{L_2 0, k_2 0}^{L'_2 0}\nonumber\\
&\times\left\lbrace
    \begin{array}{ccc}
      J_1 & k_1 & J'_1\\
      L'_1 & S & L_1
    \end{array}
    \right\rbrace
    \left\lbrace
    \begin{array}{ccc}
      J_2 & k_2 & J'_2\\
      L'_2 & S & L_2
    \end{array}\right\rbrace\nonumber\\
     &\times\sum_{p=-(k_1 + k_2)}^{k_1 + k_2} \sum_{p_1 = -k_1}^{k_1} \sum_{p_2 = -k_2}^{k_2} Y_{k_1 + k_2, p}(\hat{R})\nonumber\\
&\times C_{k_1 p_1 , k_2 p_2}^{k_1 + k_2 , p} C_{J_1 M_1, k_1 p_1}^{J'_1 M'_1} C_{J_2 M_2, k_2 p_2}^{J'_2 M'_2}.
  \label{fineangular}
\end{align}

\end{document}